%% file: main.tex
\documentclass[conference]{IEEEtran}

\usepackage[english]{babel}
\usepackage[nocompress]{cite}
\usepackage{graphicx}
\usepackage[utf8]{inputenc}
\usepackage{setspace}

\usepackage{orcidlink}

\usepackage{amsfonts}
\usepackage{amssymb}
\usepackage{caption}
\usepackage{comment}
\usepackage{multirow}
\usepackage{physics}
\usepackage{subcaption}
\usepackage{syntax}
\usepackage{url}

\usepackage{tikz}
\usetikzlibrary{quantikz2,shapes.geometric}

\newcommand{\freezecache}{frozencache,cachedir=styminted}

\usepackage[\freezecache]{minted}
\setminted{fontsize=\footnotesize}
\RecustomVerbatimEnvironment{Verbatim}{BVerbatim}{}
\newtheorem{definition}{Definition}
\newtheorem{remark}{Remark}
\newtheorem{example}{Example}

\def\eg{{\em e.g., }}
\def\ie{{\em i.e., }}
\def\silqfrag{{Silq-Hybrid}}
\def\silver{\texttt{SilVer}}
\def\silspeq{\texttt{SilSpeq}}
\def\smtlib{SMT-LIBv2}
\def\symQV{\texttt{symQV}}

\title{Automated Verification of Silq Quantum Programs using SMT Solvers}

 \author{
     \IEEEauthorblockN{Marco Lewis\IEEEauthorrefmark{1}\textsuperscript{,}\IEEEauthorrefmark{4} \orcidlink{0000-0002-4893-7658},
     Paolo Zuliani\IEEEauthorrefmark{2} \orcidlink{0000-0001-6033-5919},
     Sadegh Soudjani\IEEEauthorrefmark{3} \orcidlink{0000-0003-1922-6678}
     \\
     }
     \IEEEauthorblockA{\IEEEauthorrefmark{1}Newcastle University, Newcastle upon Tyne, UK\\}
     \IEEEauthorblockA{\IEEEauthorrefmark{2}Università degli Studi di Roma La Sapienza, Rome, Italy\\}
     \IEEEauthorblockA{\IEEEauthorrefmark{3}Max Planck Institute for Software Systems, Kaiserslautern, Germany\\}
     \IEEEauthorblockA{\IEEEauthorrefmark{4} Email: \href{mailto:m.j.lewis2@newcastle.ac.uk}{m.j.lewis2@newcastle.ac.uk}}
 }
 
\def\unary{\text{UNARY}}
\def\binary{\text{BINARY}}
\def\qcskip{\text{SKIP}}
\def\qinst{\text{QINST}}
\def\qinit{\text{QINIT}}
\def\qop{\text{QOP}}
\def\qmeas{\text{QMEAS}}

\def\op{\text{OP}}
\def\cinst{\text{CINST}}
\def\cop{\text{CSET}}
\def\cmeas{\text{CMEAS}}
\def\return{\text{RETURN}}
\def\ctrl{\text{CTRL}}
\def\ctrls{\Gamma}

\def\mem{\mathcal{M}}
\def\memc{\mem_c}
\def\memq{\mem_q}
\def\meme{\mem_e}

\def\var{x}

\def\Vars{\mathcal{X}}
\def\reg{\mathcal{R}}

\def\ff{f\!\!f}

\def\rand{\emph{rand}}
\def\cert{\emph{cert}}
\newcommand{\highprob}[1]{\emph{whp}(#1)}

\def\timeout{\tt timeout}

\begin{document}
\maketitle
\thispagestyle{plain}
\pagestyle{plain}
\input{paper}
\end{document}

%% file: paper.tex
\begin{abstract}
We present \silver{} (Silq Verification), an automated tool for verifying behaviors of quantum programs written in Silq, which is a high-level programming language for quantum computing.
The goal of the verification is to ensure correctness of the Silq quantum program against user-defined specifications using SMT solvers.
We introduce a programming model that is based on a quantum RAM-style computer as an interface between Silq programs and SMT proof obligations, allowing for control of quantum operations using both classical and quantum conditions.
Additionally, users can employ measurement \emph{flags} within the specification to easily specify conditions that measurement results require to satisfy for being a valid behavior.
We provide case studies on the verification of generating entangled states and multiple oracle-based algorithms.
\end{abstract}


\section{Introduction}
\label{sec:intro}
\input{Sections/intro}

\section{SilVer Architecture}
\input{Sections/arch}

\section{Silq and Program Verification}
\label{sec:prelim}
\input{Sections/prelim}




\section{Theoretical Foundations}
\label{sec:foundations}
\input{Sections/foundations}

\section{Verification Process}
\label{sec:verif}
\input{Sections/verif}

\section{Implementation and Case Studies}
\label{sec:casestudy}
\input{Sections/casestudy}

\section{Related Works}
\label{sec:related}
\input{Sections/relatedworks}

\section{Conclusion}
\label{sec:conclusion}
\input{Sections/conclusion}

\section*{Acknowledgements}
M.L.~was supported by the UK Engineering and Physical Sciences Research Council (EPSRC project reference EP/T517914/1). P.Z.~was supported by the project SERICS (PE00000014) under the Italian MUR National Recovery and Resilience Plan funded by the European Union - NextGenerationEU.
The research of S. Soudjani was supported by the following grants: EPSRC EP/V043676/1, EIC 101070802, and ERC 101089047.
\bibliographystyle{IEEEtran.bst}
\bibliography{IEEEabrv, refs.bib}


%% file: Sections/intro.tex
Writing quantum programs from algorithms is hard and ensuring their correctness is even tougher.
Several quantum programming languages have been released to program current or future quantum computers.
Some are libraries of classical languages, whereas others are new languages specifically designed for writing quantum programs.
Such languages include Cirq~\cite{Cirq}, Q\#~\cite{qsharp}, Qiskit~\cite{Qiskit} and Quipper~\cite{quipper}.
Silq~\cite{silq} stands out as a higher-level programming language in comparison to others, along with a typing system and semantics.
We are starting to see efforts in formally specifying languages that have no specification currently: for example, Q\# is being formalized in~\cite{Singhal22}.
Whilst some of these languages may have testing functionalities, most of them cannot be formally verified yet.
This makes implementations of quantum algorithms at risk of being incorrect.
As quantum computers become larger and programming languages become more complex, there is a need for quantum programs to be verified.

Several approaches have been taken to formally verify quantum programs, and we briefly survey some of them in Section \ref{sec:related}.
Essentially, both theorem proving and model checking approaches have been developed for verifying quantum programs.
Whilst theorem provers are powerful tools, they might require much human effort to be effective.
Thus, it is important to consider automatic techniques that offload the burden of proving a program correct to a computer.

In this paper, we present \silver{}, which is an SMT-based verification tool for reasoning about programs written in the Silq quantum programming language.
Silq programs are transformed into a symbolic structure that models a quantum RAM (QRAM)-style processor \cite{Knill96} which allows for both quantum and classical processes with a separated control channel.
Our intermediary representation, which we refer to as the QRAM Program Model, is unique as it separates out quantum and classical memories, allowing us to specify precisely the instructions affecting the memory.
Additionally, the model keeps track of controls affecting the program, which encapsulate classical conditional statements and quantum controlled operations.
\silver{} performs the conversion from Silq into the QRAM program model automatically and further converts the generated model into proof obligations.

Additionally, we provide a specification language, \silspeq{}, for reasoning about programs written in Silq.
\silspeq{} allows users to make pre-conditions about a program's inputs and post-conditions for expected results.
Importantly, pre-conditions allow for the specification of oracles for oracle-based algorithms.
A unique feature of \silspeq{} is that the user can specify properties of a measurement of a quantum state that should be obeyed using measurement \emph{flags}.
This allows a user to easily specify what properties of the measurement should be met, \eg{}the outcome of measuring a qubit in the $\ket{0}$ state should be at least 60\%.
It should be noted that conditions written in \silspeq{} are purely classical, the only interaction with quantum pre-conditions is through \emph{flags} and measured variables.

In this paper, we present \silver{} and \silspeq{}, the QRAM program model that is used to represent quantum programs, and we describe how the model is converted to SMT obligations and verified using an SMT solver, notably Z3~\cite{Z3}.
Finally, we provide case studies on programs written in Silq and verified using \silver{}.

%% file: Sections/arch.tex

\silver{} (\textbf{Sil}q \textbf{Ver}ification) is a framework for verifying Silq programs by allowing the programmer to specify behavior using a simple specification language called \silspeq{}.
The framework combines these together by converting Silq programs and \silspeq{} desired behaviors into proof obligations to be proven in a SMT solver.

\begin{figure*}[t!]
    \centering
    \input{Images/silver-tikz2}
    \caption{\silver{} Architecture}
    \label{fig:silver}
\end{figure*}

The architecture of \silver{} is given in Figure~\ref{fig:silver}.
\silver{} converts Silq programs into an intermediate representation (discussed in Section~\ref{sec:foundations:irv}), which can then be further converted into proof obligations.
These obligations are generated automatically and may change depending on the behaviors described by the \silspeq{} specification.
The behaviors that can change the generated obligations are referred to as \emph{flags} and are discussed in Section~\ref{sec:silspeq}.
Otherwise, desired behaviors in \silspeq{} can also be converted into proof obligations very easily.

The benefit of the designed \silver{} architecture is that its components are modular.
One for instance can replace Silq with another quantum programming language (such as Q\#\cite{qsharp}) and specify the behavior using \silspeq{}, which is then verified using SMT solvers.
Alternatively, one can replace the form or method of verification.
As an example, one could instead translate Silq into a theorem prover based tool (such as QBricks \cite{Chareton20} or SQIR \cite{Hietala20}), where the theorems and lemmas that are to be proved by the user are automatically generated.
The implementation details and examples of \silver{} verification are discussed in Sections~\ref{sec:casestudy}.

%% file: Images/silver-tikz2.tex
\makeatletter
\pgfdeclareshape{document}{
\inheritsavedanchors[from=rectangle] 
\inheritanchorborder[from=rectangle]
\inheritanchor[from=rectangle]{center}
\inheritanchor[from=rectangle]{north}
\inheritanchor[from=rectangle]{south}
\inheritanchor[from=rectangle]{west}
\inheritanchor[from=rectangle]{east}
\backgroundpath{
\southwest \pgf@xa=\pgf@x \pgf@ya=\pgf@y
\northeast \pgf@xb=\pgf@x \pgf@yb=\pgf@y
\pgf@xc=\pgf@xb \advance\pgf@xc by-10pt 
\pgf@yc=\pgf@yb \advance\pgf@yc by-10pt
\pgfpathmoveto{\pgfpoint{\pgf@xa}{\pgf@ya}}
\pgfpathlineto{\pgfpoint{\pgf@xa}{\pgf@yb}}
\pgfpathlineto{\pgfpoint{\pgf@xc}{\pgf@yb}}
\pgfpathlineto{\pgfpoint{\pgf@xb}{\pgf@yc}}
\pgfpathlineto{\pgfpoint{\pgf@xb}{\pgf@ya}}
\pgfpathclose
\pgfpathmoveto{\pgfpoint{\pgf@xc}{\pgf@yb}}
\pgfpathlineto{\pgfpoint{\pgf@xc}{\pgf@yc}}
\pgfpathlineto{\pgfpoint{\pgf@xb}{\pgf@yc}}
\pgfpathlineto{\pgfpoint{\pgf@xc}{\pgf@yc}}
}
}
\makeatother

\tikzstyle{doc}=[%
draw,
thick,
align=center,
color=black,
shape=document,
minimum width=20mm,
text width=17mm, 
minimum height=28.2mm,
inner sep=2ex,
]
\tikzstyle{proc}=[%
draw,
thick,
align=center,
color=black,
minimum width=20mm,
text width=17mm, 
minimum height=5mm,
shape=rectangle,
inner sep=2ex,
]

\begin{tikzpicture}[scale=0.6, every node/.style={transform shape}]
  \node[doc] (silq) {Silq\\Program};
  \node[] (space) [below = of silq] {};
  \node[doc] (silspeq) [below= of space] {SilSpeq Specification File};
  \node[proc] (SilqInterp) [right= of silq] {Silq Interpreter};
  \node[proc] (SilSpeqInterp) [right = of silspeq] {SilSpeq Interpreter};
  \node[doc,minimum height=5mm] (InternalRep) [right = of SilqInterp] {QRAM Model};
  \node[proc] (ObligationGen) [right = of InternalRep] {Obligation Generator};
  \node[doc] (ProgramObl) [right = of ObligationGen] {Program Obligations};
  \node[] (spaceObl) [below = of ProgramObl] {};
  \node[doc] (SpecObl) [below = of spaceObl] {Specification Obligations};
  \node[] (spaceSMT) [right = of spaceObl] {};
  \node[proc] (SMT) [right = of spaceSMT] {SMT Solver};

  \draw [-stealth] (silq) -- (SilqInterp);
  \draw [-stealth] (SilqInterp) -- (InternalRep);
  \draw [-stealth] (InternalRep) -- (ObligationGen);
  \draw [-stealth] (ObligationGen) -- (ProgramObl);
  \draw [-stealth] (ProgramObl) -- (SMT);

  \draw [-stealth] (silspeq) -- (SilSpeqInterp);
  \draw [-stealth] (SilSpeqInterp) -- (SpecObl);
  \draw [dashed,-stealth] (SilSpeqInterp.north east) -- (ObligationGen) node[midway,below right] {\emph{flags}};
  \draw [-stealth] (SpecObl) -- (SMT);
\end{tikzpicture}

%% file: Sections/prelim.tex
For a full introduction to quantum computing see, \eg ~\cite{NielsenChuang}. For an introduction to satisfiability see \cite{Satisfiability} and for SMT solvers see \cite{Barrett2018}, for example.

\subsection{Quantum Computing Notation}
\label{sec:bg:quantum}
\input{Sections/qc}

\subsection{Silq and the \silqfrag{} Fragment}
Silq~\cite{silq} is a high-level, imperative quantum programming language that features safe, automatic uncomputation of qubits.
In contrast, Qiskit~\cite{Qiskit} and Cirq~\cite{Cirq} are instead modules of a classical programming language.
Further, Silq features a formally defined specification; although recent work has given Q\# a formal specification \cite{Singhal22} as well.

In \cite{silq}, the authors describe a fragment of Silq (Silq-Core) for the purposes of showing its properties.
Since Silq has several high-level features, we restrict the considered programs to a fragment called \silqfrag{}.
The idea is to take a fragment of Silq that is expressive enough to handle several textbook quantum algorithms while capturing the core ideas required for verifying a program.
\silqfrag{} is based on a subset of Silq-Core and parts of Silq not contained in Silq-Core.
\begin{definition}
The expressions for {\em \silqfrag{}} are defined as follows:
\begin{equation}\label{silq-frag:expressions}
    \begin{aligned}
        e ::= &
        \; c
        \mid x
        \mid \textbf{measure}
        \mid \textbf{if } e \textbf{ then } e_1 \textbf{ else } e_2
        \\ &
        \mid \lambda(x_1, \dots, x_n). e
        \mid x := e
        \\ &
        \mid e'(e)
        \mid e'(e_1, e_2)
        \mid \textbf{return } x_c
    \end{aligned}
\end{equation}
and are
\begin{itemize}
    \item constants and built-in functions ($c$);
    \item variables ($x$);
    \item assignment ($x := e$);
    \item quantum measurement ($\textbf{measure}$);
    \item conditional statements ($\textbf{if } e \dots$);
    \item lambda abstraction ($\lambda(x_1, \dots, x_n). e$), which describes a function;
    \item function calls restricted to up to two inputs ($e'(e), e'(e_1, e_2)$);
    \item and return ($\textbf{return } x_c$) is included with the restriction that only classical variables can be returned.
\end{itemize}
\end{definition}
Importantly, conditional statements can use either quantum or classical variables within their conditions.
This allows a quantum operation to either be added or removed for a classical condition, or for a controlled quantum operation to take place if a quantum condition is used.

\silqfrag{} can represent instances of common algorithms with a specific number of qubits (\eg{} Deutsch-Jozsa~\cite{DeutschJozsa92}, GHZ state generation~\cite{Greenberger1989}, and QFT~\cite{QFT}) since it can represent quantum operations applied to qubits and represent oracles through the usage of conditional statements.
We note that it cannot represent algorithms for an arbitrary number of qubits as \silqfrag{} does not have loop expressions (\eg{} \textbf{for} and \textbf{while}).
Verifying a program using a generic number of qubits using a model checking based approach is difficult and might be better handled by theorem provers.
\begin{remark}
The Silq language itself has several syntax differences to \silqfrag{} or Silq-Core.
This is because the syntax of the Silq fragments are designed for developing theory, whereas the Silq language is designed for the programmer's convenience.
For example, conditional statements in the Silq fragments
\begin{equation*}
\textbf{if } e \textbf{ then } f \textbf{ else } g
\end{equation*}
are written in a Silq program as
\begin{equation*}
\begin{minted}[breaklines]{d}
if e { f } else { g }.
\end{minted}
\end{equation*}
As an example, Figure~\ref{fig:dj:silq} provides a 2-qubit version of the Deutsch-Jozsa algorithm.
\end{remark}

\begin{figure}[th]
    \centering
    \begin{minted}[breaklines]{d}
def fixed_dj(
f: const uint[2]!->qfree B
){
    x := 0:uint[2];
    // Apply superposition
    x[0] := H(x[0]);
    x[1] := H(x[1]);
    // Oracle
    if f(x){ phase(pi); }
    // Undo superposition
    x[0] := H(x[0]);
    x[1] := H(x[1]);
    
    x := measure(x);
    return x;
}
    \end{minted}
    \caption{Deutsch-Jozsa Silq Program for 2 qubits.
    The type of $f$ states that it is a function that takes in a quantum integer and returns a boolean.}
    \label{fig:dj:silq}
\end{figure}

\subsection{Program Verification using SMT Solvers}
\label{sec:prelim:classicalverif}
The standard technique for checking a program using SMT solvers is to convert variables into symbols that keep track of updates on a variable.
This approach is also referred to as Static Single Assignment (SSA)~\cite{Cytron86, Rosen88,Amme2001,Barthe12}.
We refer to the converted program expressions as proof obligations, and an SMT solver is called to find a model for the proof obligations.
Figure~\ref{fig:conversion:standard} shows an example of a simple classical program being converted into proof obligations and \smtlib{} format~\cite{Barrett17}.

\begin{figure}[t]
    \centering
    \begin{subfigure}[b]{0.3\textwidth}
        \centering
        \begin{minted}[autogobble]{python}
            x = 2
            x = x + 2
            y = 3 - x
        \end{minted}
        \caption{A classical program}
    \end{subfigure}
    \hfill
    \begin{subfigure}[b]{0.3\textwidth}
        \centering
        \begin{minted}[autogobble]{python}
            x0 = 2
            x1 = x0 + 2
            y0 = 3 - x1
        \end{minted}
        \caption{Proof obligations}
    \end{subfigure}
    \hfill
    \begin{subfigure}[b]{0.3\textwidth}
        \centering
        \begin{minted}[autogobble]{python}
            (= x0 2)
            (= x1 (+ x0 2))
            (= y0 (- 3 x1))
        \end{minted}
        \caption{SMT-LIB2 format}
    \end{subfigure}
    \caption{Converting a program for verification.}
    \label{fig:conversion:standard}
\end{figure}

%% file: Sections/qc.tex
We make use of the Dirac bra-ket notation to describe quantum states and operations.
The states $\ket{0} = [1, 0]^\intercal $ and $\ket{1} = [0,1]^\intercal$ describe the computational basis states for a single quantum bit (qubit).
In general, a qubit takes the form $\ket{\psi} = \beta_0 \ket{0} + \beta_1 \ket{1}$, where $\beta_i \in \mathbb{C}$ and $\abs{\beta_0}^2 + \abs{\beta_1}^2 = 1$.
To represent a system with multiple qubits, we use the tensor (or Kronecker) product to combine the quantum states, and we follow convention of not displaying its corresponding symbol unless necessary for clarity. An $n$-qubit system can be represented using the computational basis states by $\ket{\psi} = \sum_{a \in \{0,1\}^n} \beta_a \ket{a_0 a_1 \dots a_{n-1}}$, where $\beta_a \in \mathbb{C}$ and $\sum_{a \in \{0,1\}^n} \abs{\beta_a}^2 = 1$.
Note here that $a_i \in \{0,1\}$ represents the value of qubit $i$ in the state $\ket{a}$.

Unitary operations, denoted by $U$, are operations from quantum states to quantum states and their inverse is their adjoint.
Thus, $U^{-1} = U^\dagger = \overline{U^\intercal}$.
We use $U\ket{\phi}$ to mean the result of applying $U$ to $\ket{\phi}$ and we use $\otimes$ as the Kronecker product (\ie{} if $\ket{\phi} = \ket{\phi_1} \ket{\phi_2}$ and $U_i$ is a valid operation on quantum state $\ket{\phi_i}$, then $(U_1 \otimes U_2) \ket{\phi} = U_1\ket{\phi_1} U_2 \ket{\phi_2}$).
A particularly important operation is the controlled-$U$ unitary operation ($CU$).
Given a quantum system $\ket{\phi} \ket{\psi}$ where $\ket{\phi}$ is a qubit, we have $CU (\ket{0} \ket{\psi}) = \ket{0} \ket{\psi}$ and $CU (\ket{1} \ket{\psi}) = \ket{1} U\ket{\psi}$.

Measuring a single qubit $\beta_0 \ket{0} + \beta_1 \ket{1}$ in the computational basis state involves collapsing the state into one of the two computational basis states ($\ket{0}$ or $\ket{1}$).
The probability of measuring each basis state is $\abs{\beta_0}^2$ and $\abs{\beta_1}^2$, respectively.
The collapse causes the resulting basis state to be returned classically, \ie{}$\ket{0}$ returns $0$ and $\ket{1}$ returns {1}.
For an arbitrary $n$-qubit system, the measured state $\ket{a} = \ket{a_0, a_1 \dots a_{n-1}}$ is measured with probability $\abs{\beta_a}^2$.
For a quantum state $\ket{\phi}\ket{\psi} = \sum_{i=0}^{2^n-1} \beta_i \ket{i}\ket{\psi_i}$, if the result of measuring $\ket{\phi}$ is $\ket{i}$, then the quantum state after measurement is $\frac{\beta_i}{\sqrt{\abs{\beta_i}^2}}\ket{i}\ket{\psi_i}$.

%% file: Sections/foundations.tex
\subsection{Intermediate Representation for Verification}
\label{sec:foundations:irv}
As part of the conversion process, we represent Silq programs using an intermediate representation.
To handle both classical and quantum operations, we use a model based on quantum RAM (QRAM)~\cite{Knill96} to represent programs.

At a high level, our QRAM program model represents the classical and quantum memory independently.
The model receives an instruction that only affects the respective memory, \ie a quantum instruction only affects the quantum memory and a classical instruction only affects the classical memory.
The last element of the model is a list of controls, which are either classical or quantum \textbf{if} statements, that restrict the instruction being run only if the control conditions are met or, in the quantum case, enforce a controlled version of the instruction being performed.
We provide formal definitions next.

\subsubsection{Memory}
For our program, we keep track of variables through objects known as Registers, with a collection of Registers known as a Memory.
\begin{definition}
    A {\em register} is a triple $\reg = (\var, s, ver)$ where $\var \in \Vars$, $s \in \mathbb{N}$ is the number of (qu)bits required to represent $\var$ and $ver \in \mathbb{N}$ is the version number.
\end{definition}
\begin{definition}
    A {\em memory} $\mem$ is a mapping between variables and registers, where if $\var \in \Vars$ is a variable and $\reg$ is its associated register, then $\mem(\var) = \reg$.
\end{definition}
Memories are meant to be used as an interface for the obligation generator to generate appropriate obligation representations of variables.
In our implementation, one can modify a memory in a variety of ways such as adding a variable or changing the version of a register.
Registers do not keep track of the value of a variable, only properties of the variable itself (the variable symbol, its size, its version).
We track changes at the register level through $ver$ rather than the memory level as it is more efficient to track variables and only update the values of the variable when they are changed, leading to fewer proof obligations being generated.
This gives us a quantum version of Static Single Assignment (SSA).

We create two different memories for our intermediate representation: a classical memory $\memc$ and a quantum memory $\memq$.
Whilst the memories have the same functionality of keeping track of variables as the program advances, this allows us to clearly separate variables used to represent classical and quantum data.
In practice, this means how a classical memory is converted into proof obligations is different from how a quantum memory is converted.

\subsubsection{Operations}
Operations are unary or binary operators that occur between variables and/or constants.
Examples include negation ($-a$), addition ($a+b$), inequality checking ($a \leq b$) and square root (sqrt($a$)).
\begin{definition}
An {\em operation instruction}, $\op$, is defined as
\begin{equation*}
    \begin{aligned}
        \op ::= & \; \unary(\diamond, a) \mid \binary(l, \star, r),
    \end{aligned}
\end{equation*}
where $\diamond$ is some unary operation, $\star$ is a binary operation and $a, l, r$ are arguments.
\end{definition}

Operations are used to represent function application and standard built-in Silq operations ($+$, $\leq$, sqrt, \dots) applied to values or variables.
However, they do not consider how the result is used; they are given meaning through instructions and controls.

\subsubsection{Instructions}
Silq expressions are mostly modeled by instructions.
Similar to how memory is separated, we separate instructions into classical and quantum instructions because we want to distinguish when we are working on the quantum state from when we are working on classical variables.

\begin{definition}
The {\em quantum instructions}, $\qinst$, are
\begin{itemize}
    \item $\qinit(\var,n,c)$, initialization of variable $\var$ with $n$ qubits and initial state $c$;
    \item $\qop(U, \var)$, unitary evolution of a variable $\var$ using $U$;
    \item $\qmeas(\var)$, measure a quantum variable $\var$.
\end{itemize}    
\end{definition}

\begin{definition}
The classical instructions, $\cinst$, are
\begin{itemize}
    \item $\cop(\var, s, c)$, setting a variable $\var$ with $s$ bits to a value $c$;
    \item $\cop(\var, s,\op)$, setting a variable $\var$ as the result of some operation $\op$;
    \item $\cmeas(\var')$, capture the result of a measurement in a variable $\var'$;
    \item $\return(\var)$, returning a classical value $\var$.
\end{itemize}
\end{definition}

Only one of the quantum or classical instructions can be performed at a time; to accommodate this both instruction sets have access to a $\qcskip$ instruction that does nothing.
The only time a quantum and classical instruction can be performed at the same time is when a measurement is performed, which converts a quantum variable into a classical variable.

\subsubsection{Controls}
Conditional expressions are exempt from being represented as an Instruction.
The condition of a conditional expression ($\textbf{if } e \dots$) is converted into a special operation called a Control, $\ctrl$, which is simply an operation, $\op$, but with only logical operations ($\leq$, $\land$, $\lnot$, \dots).

A generated control operation is added to a vector of controls, $\ctrls$, which keeps track of all conditionals that currently affect the program.
This allows us to separate the conditions that are affecting a program from the actual instructions that are performed.
This has the additional benefit of having both quantum and classical controls inside.
This novel approach of separating controls from instructions is used to arbitrarily create controlled unitaries based on the controls used.

\subsubsection{QRAM Program Model}
When \silver{} receives a Silq program, it converts each expression in the program into a sequence of tuples\footnote{This is done by taking the abstract syntax tree of a Silq program and converting the Silq expressions into appropriate instructions (as described in Figure~\ref{fig:silver}).} of the form
\begin{equation*}
    (\qinst, \memq, \cinst, \memc, \ctrls)
\end{equation*}
where $\qinst$ ($\cinst$) is a quantum (classical) instruction, $\memq$ ($\memc$) is the quantum (classical) memory after performing the instruction and $\ctrls$ is a vector of controls ($\ctrl$).
This sequence of tuples is called a {\em QRAM Program Model}.

As described previously, we include two types of instructions that affect the respective memory and only apply one instruction in each tuple (using $\qcskip$ for the other instruction).
The only exception is when performing measurement, which uses $\qmeas$ and $\cmeas$ for $\qinst$ and $\cinst$, respectively.
This is to separate the quantum and classical parts of measurement.

The QRAM program model is usable to effectively represent any program written in the \silqfrag{} format and, therefore, most quantum circuits and some common algorithms.
We discuss SMT proof obligation generation in Section~\ref{sec:verif:convert}.

\begin{example}
    A 2-qubit version of the Deutsch-Jozsa algorithm is represented by the following QRAM program model:
    \begin{align*}
        & \big( (\qinit(x, 2, \ket{0}), \meme[x \to (x, 2, 0)] , \qcskip, \meme, \ctrls_e), \\
        & \quad (\qop(H \otimes H, x), \meme[x \to (x, 2, 1)] , \qcskip, \meme, \ctrls_e), \\
        & \quad (\qop(\pi I \otimes I, x), \meme[x \to (x, 2, 2)] , \qcskip, \meme, \ctrls_1), \\
        & \quad (\qop(H \otimes H, x), \meme[x \to (x, 2, 3)] , \qcskip, \meme, \ctrls_e), \\
        & \quad (\qmeas(x), \meme, \cmeas(x), \meme[x \to(x, 2, 4)], \ctrls_e) \big)
    \end{align*}
    where $\ctrls_e = [~]$, $\ctrls_1 = [\binary(\unary(f, x), ==, 1)]$, $\meme$ is an empty memory (for all $y$, $\meme(y) = ()$) and $\mem[x \to \reg]$ updates the memory $\mem$ such that $\mem(x) = \reg$.
\end{example}

\subsection{Behavior Specification with \silspeq{}}
\label{sec:silspeq}
\silspeq{} includes types, arithmetic and logical expressions, and measurement flags.
These are extensive enough to allow us to define useful properties for quantum algorithms as will be discussed.

\subsubsection{Variables}
In \silspeq{}, users can define variables that can be used in the pre- and post-conditions.
The types that variables can take are limited to integers of fixed bit size and functions of these types.
The types are represented by the following syntax
\begin{equation*}
    \langle type \rangle ::= \{0,1\}^n \mid \langle type \rangle \to \langle type \rangle,
\end{equation*}
where $n \in \mathbb{N}$.
A variable definition is then written as
\begin{equation*}
    \text{define } \langle name \rangle : \langle type \rangle.
\end{equation*}

\subsubsection{Expressions}
\silspeq{} uses a combination of {\em arithmetic} and {\em logical} expressions, denoted $\alpha$ and $l$ respectively, described by the grammars
\begin{equation*}
    \begin{aligned}
    \alpha ::= & \; m \in \mathbb{Z} \mid var \mid f(\alpha_1) \mid \alpha_1 * \alpha_2 \mid \alpha_1 + \alpha_2 \mid \alpha_1^{\alpha_2} \mid (\alpha_1), \\
    l ::= & \; \ff \mid \alpha_1 = \alpha_2 \mid \alpha_1 < \alpha_2 \mid \lnot l_1 \mid l_1 \& l_2 \mid @ var. l
    \end{aligned}
\end{equation*}
where $var$ is a defined variable, $f$ is a defined function, $\ff$ denotes falsity ($\bot$), $\&$ represents logical and ($\land$), and $@$ represents the universal quantifier ($\forall$).
Other arithmetic and logical expressions can be derived by combining the above (\eg $-, /$ for arithmetic expressions and $\leq, >, \lor, \implies, \exists, \dots$ for logical expressions).

\subsubsection{Flags}
Many quantum verification frameworks branch on measurement; \ie if a qubit is measured and the result is $\ket{1}$, perform a command, otherwise perform another.
In Silq-Hybrid, this will be in the form $\textbf{if measure } x \textbf{ then } e_1 \textbf{ else } e_2$.
In \silspeq{}, we instead consider the properties of measurement that we wish to verify as well.
Specifically, we include \emph{flags} which tell us what measurement property we wish to verify for a given function.
This changes the measurement outcome that can occur.
Additionally, this allows us to restrict the cases we need to consider.
For instance, if we allow only measurement to be done with certainty (the register state being measured is $\beta\ket{b}$ for $\abs{\beta} = 1, b \in \{0,1\}^n$), then the solver only needs to consider measurement values that satisfy such constraint.
On the other hand, the flexibility of flags mean that the solver can consider all possible measurement values and take these into account as well.

Flags mainly affect measurement and interact with the obligations generated from measurement instructions ($\qmeas$ and $\cmeas$).
For numerous quantum algorithms, there are certain properties about measurement that we want to be able to verify.
The types of flags available are:
\begin{itemize}
    \item The \rand{} flag puts no obligations on measurement and so any measurement result is possible so long as it has a non-zero likelihood of occurring.
    \item The \cert{} flag specifies that for any run of the program, there must be a state that has a measurement probability of $1$, \ie with certainty.
    \item The $\highprob{x}$ (with high probability) flag states that for any program run, we only consider measurement results that occur with probability greater than $x$.
    For ease, we write $\emph{whp} = \highprob{0.5}$ and note that $\cert{} = \highprob{1}$.
\end{itemize}

\subsubsection{Discussion}
\silspeq{} is used for specifying the behavior of programs that have purely classical input, quantum operations and measurement during program execution, and a classical result.
We discuss some of the features of \silspeq{} that both benefit and limit its scope:
\begin{itemize}
    \item Function behaviors for oracles can be specified in a concise way.
    For example, the oracles for the Deutsch-Jozsa and the Bernstein-Vazirani algorithms are described in the next section for a 2 qubit program (given in Figures~\ref{fig:dj:silspeq} and \ref{fig:bv}).
    \item \silspeq{} considers only classical input and output.
    If the function to verify is a quantum operation, such as preparing a target state ($\ket{\phi}$), then \silspeq{} cannot specify if the output of a program matches the target state.
    \item \silspeq{} only allows specification of quantum properties through flags.
    This allows the users to specify what properties of measurement they want without needing to define a measurement operator matrix.
    \item One cannot perform different flags for different measure instructions.
    Programs are assumed to have a single measurement or measurements that require the same property.
    \item Temporal properties cannot be specified.
    Neither quantum nor classical variables may be used to specify behavior at given points in the program execution.
    This allows \silver{} to automate the conversion of the input program.
\end{itemize}
We provide some information on how to read \silspeq{} specification in Figure~\ref{fig:silspeqread} and examples of specifications in Figure~\ref{fig:silspeqex}.
\begin{figure}[t]
    \centering
    \begin{minted}[breaklines]{text}
prog[flags](input : t1)->
    (define prog_ret : t2)
pre{
    define x : t3
    assert(input - x <= x)
}
post{
    assert(prog_ret = x)
}
    \end{minted}
    \caption{Reading \silspeq{} specification: \texttt{prog} is the name of the function to be verified, \texttt{input} is a function input with type \texttt{t1}, \texttt{prog\_ret} is the output of a function with type \texttt{t2}, \texttt{x} is a defined variable with type \texttt{t3} and we have several \texttt{assert} statements containing logical expressions.}
    \label{fig:silspeqread}
\end{figure}

\begin{figure}[t]
    \centering
\begin{subfigure}[b]{\columnwidth}
    \centering
    \begin{minted}[breaklines]{text}
unfair_coin[whp(0.75)]()->
    (define unfair_coin_ret : {0,1})
pre{
}
post{
    assert(prog_ret = 0)
}
    \end{minted}
    \caption{\silspeq{} for specification of an unfair quantum coin: at least 75\% return 0, otherwise return 1.}
    \label{fig:unfaircoin}
    \vspace{.25cm}
\end{subfigure}
\begin{subfigure}[b]{\columnwidth}
    \centering
    \begin{minted}[breaklines]{text}
multiple_5[rand]()->
    (define multiple_5_ret : {0,1}^5)
pre{
}
post{
    assert(multiple_5_ret % 5 = 0)
}
    \end{minted}
    \caption{\silspeq{} for a program that returns a multiple of 5, no matter the result of any quantum measurement.}
    \label{fig:multiple5}
    \vspace{.25cm}
\end{subfigure}
\begin{subfigure}[b]{\columnwidth}
    \centering
    \begin{minted}[breaklines]{text}
always_1[cert]()->
    (define always_1_ret : {0,1}^2)
pre{
}
post{
    assert(always_1_ret = 1)
}
    \end{minted}
    \caption{\silspeq{} for a program that always returns 1 and any quantum measurement is performed with certainty.}
    \label{fig:always1}
\end{subfigure}
\caption{\silspeq{} examples}
\label{fig:silspeqex}
\end{figure}

%% file: Sections/verif.tex
For Silq programs and \silspeq{} specifications, \silver{} converts them to SMT encodings.
A Silq program is converted into the QRAM program model first, and then into \smtlib{} format, which we denote by $prog$.
\silspeq{} specifications can be directly converted into an SMT encoding: $pre$ and $post$ represent a \silspeq{} specification's pre- and post-conditions SMT encoding respectively.

\subsection{Conversion of QRAM Program Model to SMT Format}
\label{sec:verif:convert}
The creation of $pre$ and $post$ simply entails converting the logical and arithmetic expressions into the appropriate proof obligations.
The number of obligations generated remains about the same amount as the number of assertions given.

\silqfrag{} programs are converted to {quantifier-free} {non-linear} real arithmetic (QF\_NRA) proof obligations through conversion of the QRAM program model.
Classical memory is used to create proof obligations using SSA form in the usual way.
The quantum memory is converted into an SSA form of the quantum state, which is a vector of complex variables that can be represented in SMT format by using two real variables.
For example, a single qubit $q$ is represented by $[q0_r + i q0_i, q1_r + i q1_i]$ where $qb_r$ and $qb_i$ are real variables representing the real and complex part of the amplitude in state $\ket{b}$ respectively.

As a reminder, the main purpose of this work is to explore the usage of an automated software verification framework within the context of quantum programs and the challenges that are faced.
Whilst using a complex vector is not an efficient way to represent the quantum state, it allows us to represent any quantum state exactly and is simple to implement.
More efficient quantum state representations include tensor networks \cite{biamonte2017tensor} and matrix product states \cite{PerecGarcia07}, although their worst-case complexity remains exponential in the number of qubits.
Another way to represent {\em single} quantum bits exactly is the Bloch sphere used by \symQV{}~\cite{symQV}.
To reduce the complexity one could recognize if a quantum program only uses certain gates that admit efficient simulation (\eg the Clifford set \cite{StabSim}).
Also, even identifying if a quantum program can represent the quantum state using only real numbers with no complex component would half the number of obligations generated for the quantum part of a program.

\subsection{Verification of Encodings}
When a program is checked against its specification, the logical statement
\begin{equation}\label{eq:SMTcheck}
    prog \land pre \land \lnot post
\end{equation}
is checked by an SMT solver, which looks for an execution (a model from $prog$), with pre-conditions, given by $pre$, that does not satisfy the post-conditions, $post$, given by the user.
If the conjunction in Equation~\eqref{eq:SMTcheck} is satisfiable, then we have a program execution that does not meet the user defined specification.
However, if the conjunction is unsatisfiable and we show that $prog \land pre$ is satisfiable, then all possible executions adhere to the specification and end in the state $post$.

%% file: Sections/casestudy.tex
\silver{} is implemented in Python with over 2{,}000 lines of code, 300 of which are used for implementing \silspeq{}.
The Python interface for Z3~\cite{Z3}, z3py, is used to generate obligations.
Users provide Silq programs, restricted to the \silqfrag{} format, and \silspeq{} specification to be verified.

Our benchmarks are tested on specified qubit instances of the following programs:
GHZ state generation,
the Deutsch-Jozsa algorithm,
and the Bernstein-Vazirani algorithm.
The provided Silq programs are based on test files found in the Silq repository.\footnote{\url{https://github.com/eth-sri/silq/tree/master/test} (accessed 22/03/2024)}

\subsection{GHZ State Generation}
The Greenberger–Horne–Zeilinger (GHZ) state~\cite{Greenberger1989} for $n$ qubits is
$
(\frac{1}{\sqrt{2}} \ket{0}^{\otimes{n}} + \frac{1}{\sqrt{2}} \ket{1}^{\otimes{n}}),
$
\ie we have a 50/50 chance of reading $n$ 0's or $n$ 1's.
Figure~\ref{fig:ghz} shows the program and the specification for setting up the 2-qubit GHZ state (also known as a Bell state) in the $y$ register using $x$ as an additional qubit.
The $\highprob{}$ flag is used to specify that we should only measure one of two possible results with 50\% probability, and that no preconditions are required.
Both program and specification can be easily extended to any qubit number.
\begin{figure}[t!]
    \centering
    \begin{subfigure}[b]{\columnwidth}
    \centering
    \begin{minted}{d}
def ghz(){
    x := 0:B;
    y := 0:uint[2];
    // Apply superposition
    x := H(x);
    // Apply controlled-NOTs
    if x{
        y[0] := X(y[0]);
        y[1] := X(y[1]);
    }
    y := measure(y);
    x := measure(x);
    return y;
}
    \end{minted}
    \caption{GHZ Silq Program}
    \vspace{.25cm}
    \end{subfigure}
    \begin{subfigure}[b]{\columnwidth}
    \centering
    \begin{minted}[breaklines]{text}
ghz[whp(0.5)]()->
    (define ghz_ret:{0, 1}^2)
pre{
}
post{
    assert(ghz_ret = 0 |
        ghz_ret = 3)
}
    \end{minted}
    \caption{GHZ \silspeq{} Specification}
    \end{subfigure}
    \caption{Generation of GHZ states in Silq (a); and correctness specification (b).}
    \label{fig:ghz}
\end{figure}
\subsection{Deutsch-Jozsa Algorithm}
This algorithm~\cite{DeutschJozsa92} was one of the first oracle-based quantum algorithms to be developed, and can distinguish if a function has one of two properties. A function $f: X \to \{0,1\}$ is said to be \emph{constant} if all inputs return the same value ($0$ or $1$).
Further, $f$ is \emph{balanced} if half the inputs of $f$ return $0$ and the other half return $1$.
Given $f : \{0,1\}^n \to \{0,1\}$ which is either constant or balanced, the Deutsch-Jozsa algorithm correctly decides which property $f$ has, using only a single evaluation of $f$. The algorithm returns $0$ if the function is constant, otherwise it returns a non-zero value.
The Silq program and \silspeq{} specification for a two qubit version of the Deutsch-Jozsa algorithm are given in Figures~\ref{fig:dj:silq} and \ref{fig:dj:silspeq} respectively (they both can be easily extended to any qubit number).

\begin{figure}
    \centering
    \begin{minted}[breaklines]{text}
fixed_dj[rand](define f:{0, 1}^2->{0, 1})->
    (define fixed_dj_ret : {0, 1}^2)
pre{
    define y : N
    define x : {0,1}^2
    define bal : {0,1}
    assert(SUM[x](f) = y)
    assert((bal = 0 & (y = 0 | y = 4))
        | (bal = 1 & y = 2))
}
post{
    assert(bal = 0 -> fixed_dj_ret = 0)
    assert(bal = 1 -> ¬fixed_dj_ret = 0)
}
    \end{minted}
    \caption{Deutsch-Jozsa \silspeq{} specification for 2 qubits.
    In the specification, $\text{SUM}[x](f) \equiv \sum_{x=0}^3f(x)$.
    }
    \label{fig:dj:silspeq}
\end{figure}
\subsection{Bernstein-Vazirani Algorithm}
The problem the Bernstein-Vazirani algorithm \cite{BernsteinVazirani97} solves is: given an oracle implementing $f : \{0,1\}^n \to \{0,1\}$ such that $f(x) = s.x$ for some $s \in \{0,1\}^n$, find $s$.
The dot product is $s.x = s_0 x_0 \oplus \dots \oplus s_{n-1} x_{n-1}$ where $\oplus$ denotes addition modulo 2.
The algorithm is the same as the Deutsch-Jozsa.
The 2-qubit Silq program for the Bernstein-Vazirani algorithm is the same program used for the Deutsch-Jozsa algorithm, given in Figure~\ref{fig:dj:silq}.
Specification for the behavior of the Bernstein-Vazirani algorithm on two qubits is given in Figure~\ref{fig:bv}.
Again, both the Silq program and its \silspeq{} specification can be easily extended to any size.
\begin{figure}[t!]
    \centering
    \begin{minted}[breaklines]{text}
fixed_bernvaz[cert](define f:{0, 1}^2->{0, 1})->
    (define fixed_bernvaz_ret : {0, 1}^2)
pre{
    define s : {0,1}^2
    define x : {0,1}^2
    assert(@x. f(x) = (s.x) mod 2)
}
post{
    assert(fixed_bernvaz_ret = s)
}
    \end{minted}
    \caption{\silspeq{} specification for the Bernstein-Vazirani algorithm.
    Note that $@x. \phi(x)$ denotes $\forall x. \phi(x)$ and $s.x$ is the dot product as described in Section~\ref{sec:casestudy}.
    }
    \label{fig:bv}
\end{figure}

\subsection{Benchmarks}
\begin{table*}[t]
    \centering
    \caption{
    Benchmarks of running \silver{} on different programs:
    Qubits = number of qubits used;
    Setup Time = CPU time for \silver{} to translate a Silq program and \silspeq{} obligations into the SMT solver;
    Verification Time = CPU time for the solver to check the obligations are unsatisfiable;
    Memory = RAM used by the SMT solver. Times are average and standard deviation over 10 runs. The timeout is set to 3 hours.
    }
    \setlength{\tabcolsep}{0.3em}
    \small{
    \begin{tabular}{|c|c|c|c|c|c|}
         \hline
         Benchmark & Qubits & Setup Time (s) & Verification Time (s) & Memory (MiB)\\
         \hline
         \multirow{4}{*}{GHZ}& 2 & 0.14$\pm 0.02$ & 0.003$\pm 5 \times10^{-4}$ & 17\\
         & 5 & 3.4$\pm 0.03$ & 0.029$\pm 9\times 10^{-4}$& 18 \\
         & 7 & 53$\pm 1.8$ & 0.4$\pm 0.02$ & 95 \\
         & 8 & 210$\pm 3.9$ & 1.7$\pm 0.07$ & 350 \\
         \hline
         \multirow{4}{*}{\shortstack{Deutsch-\\Jozsa}}& 2 & 0.066$\pm0.019$ & 0.009$\pm0.002$ & 17\\
         & 3 & 0.15$\pm0.02$ & 0.041$\pm0.007$ & 17\\
         & 4 & 0.41$\pm0.05$ & 7.9$\pm1$ & 19\\
         & 5 & 1.4$\pm0.1$ & \timeout  & 23\\
         \hline
         \multirow{6}{*}{\shortstack{Bernstein-\\Vazirani}}& 2 & 0.06$\pm 0.01$ & 0.01$\pm0.001$ & 17 \\
         & 3 & 0.14$\pm 0.01$ & 0.07$\pm 0.01$ & 17 \\
         & 4 & 0.46$\pm 0.06$ & 0.5$\pm 0.03$ & 18 \\
         & 5 & 3.8$\pm2.2 $ & 3.7$\pm 0.36$ & 19 \\
         & 6 & 13$\pm 1.3$ & 41$\pm 7.1$ & 39 \\
         & 7 & 30$\pm 0.46$ & 1000$\pm 260$ & 67 \\
         \hline
    \end{tabular}
    }
    \label{tab:benchmarks}
\end{table*}

The experiments were performed on a laptop with an Intel(R) Core(TM) i5-10310U CPU @ 1.70GHz x 8 cores processor and 16GB of RAM, running Ubuntu 20.04.3 LTS.
The results are the average of 10 runs (with standard deviation) and are given in Table~\ref{tab:benchmarks}.
We do not compare our tool against theorem proving tools~\cite{Hietala20,Chareton20,Liu19, CoqQ}, since these are manual approaches to verifying programs.
Additionally, we do not compare with \symQV{}~\cite{symQV} since the type of specification checked between each tool is different.

\subsection{Discussion}
The performance of \silver{} depends on the input program's complexity, which can depend on the number of oracle functions to consider.
In some cases verification setup takes a much longer time than actual verification (\eg GHZ) whereas verification becomes slower in other cases (\eg Bernstein-Vazirani).
The setup time grows exponentially based on the number of qubits due to the eventual usage of matrices during the setup process.
Since \silver{} uses a basic approach to performing the verification (through standard complex vector representation), these times represent targets that other techniques would need to beat both in terms of transforming a program into a verifiable format and the verification time itself.

With regards to algorithms, checking the Bernstein-Vazirani algorithm becomes faster than checking Deutsch-Jozsa, after a certain qubit size.
That is because the state space for the latter grows much faster than for the former. 
Specifically, for $n$ qubit problems, there are $2^n$ possible functions for Bernstein-Vazirani (since the secret string $s \in \{0,1\}^n$), whereas for Deutsch-Jozsa there are ${\binom{2^n}{2^{n-1}} + 2} \approx \frac{2^{2^n}}{\sqrt{2^n}}$ possible functions (due to the nature of balanced functions).
This is where theorem provers may be a better tool for verification of certain quantum algorithms.
This also explains why checking the GHZ states is quick, since it only needs to check that the program transforms the initial state, $\ket{0}$, into the desired GHZ state.

There are some restrictions to \silver{}.
If running on a standard laptop, \silver{} will unlikely verify anything larger than 10 qubits.
Additionally, the \smtlib{} files for the obligations become large very quickly; a 3-qubit instance of the Bernstein-Vazirani algorithm only requires about 20kB of space whereas the 10-qubit version takes up over 250MB of memory.

%% file: Sections/relatedworks.tex
Theorem proving approaches to quantum program verification include SQIR~\cite{Hietala20}, QHLProver~\cite{Liu19} and CoqQ~\cite{CoqQ}.
Model checking-based approaches such as QPMC~\cite{QPMC} and Entang$\lambda$e~\cite{Anticoli18} allow users to convert programs into quantum Markov chains, which are then checked against a temporal logic specification.
QBricks~\cite{Chareton20} makes use of the Why3 framework to prove properties about programs or provide a counterexample to the property specified.
However, QBricks requires the user to learn a domain-specific language and to write their program within the Why3 framework.
Giallar~\cite{Giallar} and the quantum program model for \symQV{}~\cite{symQV} use SMT solvers to perform verification on the Qiskit compiler (rather than on input quantum programs).
Recent approaches for verifying quantum circuits include combining decision diagrams with symbolic verification~\cite{hong2023decision} and using automata-based reasoning~\cite{Chen23a, chen23b}.
For more detailed surveys on formal verification for quantum computing see \cite{Chareton23, lewis2022formal}.


We note that SQIR~\cite{Hietala20}, \symQV{}~\cite{symQV}, and QBricks~\cite{Chareton20} also use intermediate representations.
However, our representation acts as a model for a QRAM device~\cite{Knill96}, where a classical computer sends instructions to a quantum chip to execute.
Apart from the model used by QBricks, other models focus purely on quantum circuits rather than instructions for QRAM devices, which are more complex and allow for classical operations to occur.
Our approach not only has quantum controls, but classical controls; meaning uncontrolled unitary operations are considered in the quantum instructions.
In addition, one can translate a quantum programming language, such as in this paper with Silq, into a general representation that can then be translated into different formats for different purposes of verification.
Whilst this modularity is achievable with other models, the models focus on quantum circuits rather than quantum programs.

Whilst \silver{} uses flags to handle measurement of quantum variables, other approaches exist.
In \cite{Singhal2022separation}, the authors use the idea of ghost variables to keep track of partially measured and entangled states.
Logical simplifications rules can then be used to remove the ghost variables to find out the likelihood of certain behaviors.
Essentially, ghost variables allow for the measurement of quantum variables to be delayed until simplification rules can be applied.
This is similar to how flags work in \silver{}, where only certain behaviors are achieved with a certain likelihood.
However, \silver{} leaves the handling of the measurement to the SMT solver.
Using ghost variables in an automated or manual verification tools needs to be implemented and seen in practice.

Another specification language is ScaffML~\cite{ScaffML}, a behavioral interface specification language (BISL) for the Scaffold quantum programming language~\cite{Scaffold}.
ScaffML uses measurement in the specification to determine how a gate or module behaves based upon the measurement outcome.
While ScaffML is capable of specifying how quantum gates behave, a feature not in \silver{}, it is done through the usage of computational basis states.
This works well for small quantum gates, but requires generalization to handle programs or general unitary operations.
\silver{} is partly capable of this, however this is through the usage of measurement operations.
Combining some of the ideas in \silver{} and ScaffML could prove useful for specifying general quantum operations.

The modular design of \silver{} is influenced by the LLVM-based intermediate program representation Vellvm~\cite{Vellvm}.
Similarly, the quantum intermediate representation (QIR) is a LLVM-based representation for quantum programming languages~\cite{QIRAlliance}.
Work has already been done on formalizing QIR code safety~\cite{Luo23}.
%

%% file: Sections/conclusion.tex
We presented \silver{}, a tool for automatically verifying programs written in the Silq quantum programming language.
\silver{} aims at making the verification of programs as simple and automatic as possible.
This includes introducing {flags} for ensuring certain properties from measurement and the automatic translation of Silq programs into a model that is converted into proof obligations. Future work includes expanding the features of Silq supported for verification, adapting the Vellvm approach for the QIR, and exploring symbolic approaches based on efficient data structures such as decision diagrams.

\paragraph*{Data Availability}
An artifact containing Silq, \silver{} and the benchmark experiments is available \cite{artifact}.

%% file: main.bbl
\begin{thebibliography}{10}
\providecommand{\url}[1]{#1}
\csname url@samestyle\endcsname
\providecommand{\newblock}{\relax}
\providecommand{\bibinfo}[2]{#2}
\providecommand{\BIBentrySTDinterwordspacing}{\spaceskip=0pt\relax}
\providecommand{\BIBentryALTinterwordstretchfactor}{4}
\providecommand{\BIBentryALTinterwordspacing}{\spaceskip=\fontdimen2\font plus
\BIBentryALTinterwordstretchfactor\fontdimen3\font minus \fontdimen4\font\relax}
\providecommand{\BIBforeignlanguage}[2]{{%
\expandafter\ifx\csname l@#1\endcsname\relax
\typeout{** WARNING: IEEEtran.bst: No hyphenation pattern has been}%
\typeout{** loaded for the language `#1'. Using the pattern for}%
\typeout{** the default language instead.}%
\else
\language=\csname l@#1\endcsname
\fi
#2}}
\providecommand{\BIBdecl}{\relax}
\BIBdecl

\bibitem{Cirq}
{Cirq Developers}, ``Cirq,'' Dec. 2022, {See full list of authors on Github: \url{https://github.com/quantumlib/Cirq/graphs/contributors}}.

\bibitem{qsharp}
K.~Svore, A.~Geller, M.~Troyer, J.~Azariah, C.~Granade, B.~Heim, V.~Kliuchnikov, M.~Mykhailova, A.~Paz, and M.~Roetteler, ``Q\#: Enabling scalable quantum computing and development with a high-level {DSL},'' in \emph{Proceedings of the Real World Domain Specific Languages Workshop 2018}, ser. RWDSL2018.\hskip 1em plus 0.5em minus 0.4em\relax ACM, 2018.

\bibitem{Qiskit}
{Qiskit contributors}, ``Qiskit: An open-source framework for quantum computing,'' 2023.

\bibitem{quipper}
A.~S. Green, P.~L. Lumsdaine, N.~J. Ross, P.~Selinger, and B.~Valiron, ``Quipper: A scalable quantum programming language,'' \emph{SIGPLAN Not.}, vol.~48, no.~6, p. 333–342, Jun. 2013.

\bibitem{silq}
B.~Bichsel, M.~Baader, T.~Gehr, and M.~Vechev, ``Silq: A high-level quantum language with safe uncomputation and intuitive semantics,'' in \emph{Proceedings of the 41st ACM SIGPLAN Conference on Programming Language Design and Implementation}, ser. PLDI 2020.\hskip 1em plus 0.5em minus 0.4em\relax New York, NY, USA: Association for Computing Machinery, 2020, p. 286–300.

\bibitem{Singhal22}
K.~Singhal, K.~Hietala, S.~Marshall, and R.~Rand, ``Q\# as a quantum algorithmic language,'' in \emph{Proceedings of the 19th International Conference on Quantum Physics and Logic (QPL), Oxford, U.K., June 27--July 1, 2022}, June 2022.

\bibitem{Knill96}
\BIBentryALTinterwordspacing
E.~Knill, ``Conventions for quantum pseudocode,'' 1996. [Online]. Available: \url{https://www.osti.gov/biblio/366453}
\BIBentrySTDinterwordspacing

\bibitem{Z3}
L.~{de}~Moura and N.~Bj\o{}rner, ``{Z3: An Efficient SMT Solver},'' in \emph{TACAS 2008}, ser. LNCS, vol. 4963.\hskip 1em plus 0.5em minus 0.4em\relax Springer-Verlag, 2008, p. 337–340.

\bibitem{Chareton20}
C.~Chareton, S.~Bardin, F.~Bobot, V.~Perrelle, and B.~Valiron, ``An automated deductive verification framework for circuit-building quantum programs,'' in \emph{ESOP 2021}, N.~Yoshida, Ed.\hskip 1em plus 0.5em minus 0.4em\relax Springer, 2021, pp. 148--177.

\bibitem{Hietala20}
K.~Hietala, R.~Rand, S.-H. Hung, L.~Li, and M.~Hicks, ``{Proving Quantum Programs Correct},'' in \emph{12th International Conference on Interactive Theorem Proving (ITP 2021)}, ser. Leibniz International Proceedings in Informatics (LIPIcs), L.~Cohen and C.~Kaliszyk, Eds., vol. 193.\hskip 1em plus 0.5em minus 0.4em\relax Dagstuhl, Germany: Schloss Dagstuhl -- Leibniz-Zentrum f{\"u}r Informatik, 2021, pp. 21:1--21:19.

\bibitem{NielsenChuang}
M.~A. Nielsen and I.~L. Chuang, \emph{Quantum Computation and Quantum Information (10th Anniversary edition)}.\hskip 1em plus 0.5em minus 0.4em\relax Cambridge University Press, 2016.

\bibitem{Satisfiability}
\BIBentryALTinterwordspacing
A.~Biere, M.~Heule, H.~van Maaren, and T.~Walsh, Eds., \emph{Handbook of Satisfiability}, ser. Frontiers in Artificial Intelligence and Applications, vol. 185.\hskip 1em plus 0.5em minus 0.4em\relax IOS Press, 2009. [Online]. Available: \url{http://dblp.uni-trier.de/db/series/faia/faia185.html}
\BIBentrySTDinterwordspacing

\bibitem{Barrett2018}
C.~Barrett and C.~Tinelli, ``Satisfiability modulo theories,'' in \emph{Handbook of Model Checking}, E.~M. Clarke, T.~A. Henzinger, H.~Veith, and R.~Bloem, Eds.\hskip 1em plus 0.5em minus 0.4em\relax Springer, 2018, pp. 305--343.

\bibitem{DeutschJozsa92}
D.~Deutsch and R.~Jozsa, ``Rapid solution of problems by quantum computation,'' \emph{Proceedings of the Royal Society of London. Series A: Mathematical and Physical Sciences}, vol. 439, no. 1907, pp. 553--558, 1992.

\bibitem{Greenberger1989}
D.~M. Greenberger, M.~A. Horne, and A.~Zeilinger, ``Going beyond bell's theorem,'' in \emph{Bell's Theorem, Quantum Theory and Conceptions of the Universe}, M.~Kafatos, Ed.\hskip 1em plus 0.5em minus 0.4em\relax Springer Netherlands, 1989, pp. 69--72.

\bibitem{QFT}
\BIBentryALTinterwordspacing
D.~Coppersmith, ``An approximate fourier transform useful in quantum factoring,'' 2002. [Online]. Available: \url{arxiv:quant-ph/0201067}
\BIBentrySTDinterwordspacing

\bibitem{Cytron86}
R.~Cytron, A.~Lowry, and F.~K. Zadeck, ``Code motion of control structures in high-level languages,'' in \emph{Proceedings of the 13th ACM SIGACT-SIGPLAN Symposium on Principles of Programming Languages}, ser. POPL '86.\hskip 1em plus 0.5em minus 0.4em\relax ACM, 1986, p. 70–85.

\bibitem{Rosen88}
B.~K. Rosen, M.~N. Wegman, and F.~K. Zadeck, ``Global value numbers and redundant computations,'' in \emph{Proceedings of the 15th ACM SIGPLAN-SIGACT Symposium on Principles of Programming Languages}, ser. POPL '88.\hskip 1em plus 0.5em minus 0.4em\relax ACM, 1988, p. 12–27.

\bibitem{Amme2001}
W.~Amme, N.~Dalton, J.~von Ronne, and M.~Franz, ``Safetsa: A type safe and referentially secure mobile-code representation based on static single assignment form,'' \emph{SIGPLAN Not.}, vol.~36, no.~5, p. 137–147, May 2001.

\bibitem{Barthe12}
G.~Barthe, D.~Demange, and D.~Pichardie, ``{A Formally Verified SSA-Based Middle-End},'' in \emph{ESOP 2012}, ser. LNCS, H.~Seidl, Ed., vol. 7211.\hskip 1em plus 0.5em minus 0.4em\relax Springer, 2012, pp. 47--66.

\bibitem{Barrett17}
C.~Barrett, P.~Fontaine, and C.~Tinelli, ``{The SMT-LIB Standard: Version 2.6},'' Department of Computer Science, The University of Iowa, Tech. Rep., 2017, available at \url{www.SMT-LIB.org}.

\bibitem{biamonte2017tensor}
\BIBentryALTinterwordspacing
J.~Biamonte and V.~Bergholm, ``Tensor networks in a nutshell,'' 2017. [Online]. Available: \url{arXiv:1708.00006}
\BIBentrySTDinterwordspacing

\bibitem{PerecGarcia07}
D.~Perez-Garcia, F.~Verstraete, M.~M. Wolf, and J.~I. Cirac, ``Matrix product state representations,'' \emph{Quantum Info. Comput.}, vol.~7, no.~5, p. 401–430, Jul. 2007.

\bibitem{symQV}
F.~Bauer-Marquart, S.~Leue, and C.~Schilling, ``sym{QV}: Automated symbolic verification of quantum programs,'' in \emph{FM 2023}, M.~Chechik, J.-P. Katoen, and M.~Leucker, Eds.\hskip 1em plus 0.5em minus 0.4em\relax Springer, 2023, pp. 181--198.

\bibitem{StabSim}
S.~Aaronson and D.~Gottesman, ``Improved simulation of stabilizer circuits,'' \emph{Phys. Rev. A}, vol.~70, p. 052328, 2004.

\bibitem{BernsteinVazirani97}
E.~Bernstein and U.~Vazirani, ``Quantum complexity theory,'' \emph{SIAM Journal on Computing}, vol.~26, no.~5, pp. 1411--1473, 1997.

\bibitem{Liu19}
J.~Liu, B.~Zhan, S.~Wang, S.~Ying, T.~Liu, Y.~Li, M.~Ying, and N.~Zhan, ``Formal verification of quantum algorithms using quantum {H}oare logic,'' in \emph{CAV 2019}, ser. LNCS, I.~Dillig and S.~Tasiran, Eds., vol. 11562.\hskip 1em plus 0.5em minus 0.4em\relax Springer, 2019, pp. 187--207.

\bibitem{CoqQ}
L.~Zhou, G.~Barthe, P.-Y. Strub, J.~Liu, and M.~Ying, ``{CoqQ}: Foundational verification of quantum programs,'' \emph{Proc. ACM Program. Lang.}, vol.~7, no. POPL, 2023.

\bibitem{QPMC}
Y.~Feng, E.~M. Hahn, A.~Turrini, and L.~Zhang, ``{QPMC}: A model checker for quantum programs and protocols,'' in \emph{FM 2015}, ser. LNCS, N.~Bj{\o}rner and F.~de~Boer, Eds., vol. 9109.\hskip 1em plus 0.5em minus 0.4em\relax Springer, 2015, pp. 265--272.

\bibitem{Anticoli18}
L.~Anticoli, C.~Piazza, L.~Taglialegne, and P.~Zuliani, ``$\mathtt{Entang{\lambda}e}$: A translation framework from quipper programs to quantum {M}arkov chains,'' in \emph{New Frontiers in Quantitative Methods in Informatics}, S.~Balsamo, A.~Marin, and E.~Vicario, Eds.\hskip 1em plus 0.5em minus 0.4em\relax Springer, 2018, pp. 113--126.

\bibitem{Giallar}
R.~Tao, Y.~Shi, J.~Yao, X.~Li, A.~Javadi-Abhari, A.~W. Cross, F.~T. Chong, and R.~Gu, ``Giallar: Push-button verification for the {Q}iskit quantum compiler,'' in \emph{Proceedings of the 43rd ACM SIGPLAN International Conference on Programming Language Design and Implementation}, ser. PLDI 2022.\hskip 1em plus 0.5em minus 0.4em\relax ACM, 2022, p. 641–656.

\bibitem{hong2023decision}
\BIBentryALTinterwordspacing
X.~Hong, W.-J. Huang, W.-C. Chien, Y.~Feng, M.-H. Hsieh, S.~Li, C.-S. Yeh, and M.~Ying, ``Decision diagrams for symbolic verification of quantum circuits,'' 2023. [Online]. Available: \url{arXiv:2308.00440}
\BIBentrySTDinterwordspacing

\bibitem{Chen23a}
Y.-F. Chen, K.-M. Chung, O.~Leng{\'a}l, J.-A. Lin, and W.-L. Tsai, ``{AUTOQ}: An automata-based quantum circuit verifier,'' in \emph{Computer Aided Verification}, C.~Enea and A.~Lal, Eds.\hskip 1em plus 0.5em minus 0.4em\relax Springer Nature Switzerland, 2023, pp. 139--153.

\bibitem{chen23b}
Y.-F. Chen, K.-M. Chung, O.~Leng\'{a}l, J.-A. Lin, W.-L. Tsai, and D.-D. Yen, ``An automata-based framework for verification and bug hunting in quantum circuits,'' \emph{Proc. ACM Program. Lang.}, vol.~7, no. PLDI, Jun. 2023.

\bibitem{Chareton23}
C.~Chareton, S.~Bardin, D.~Lee, B.~Valiron, R.~Vilmart, and Z.~Xu, ``Formal methods for quantum algorithms,'' in \emph{Handbook of Formal Analysis and Verification in Cryptography}, 1st~ed., S.~Akleylek and B.~Dundua, Eds.\hskip 1em plus 0.5em minus 0.4em\relax {CRC} Press, 2023.

\bibitem{lewis2022formal}
\BIBentryALTinterwordspacing
M.~Lewis, S.~Soudjani, and P.~Zuliani, ``Formal verification of quantum programs: Theory, tools, and challenges,'' \emph{ACM Transactions on Quantum Computing}, vol.~5, no.~1, Dec. 2023. [Online]. Available: \url{https://doi.org/10.1145/3624483}
\BIBentrySTDinterwordspacing

\bibitem{Singhal2022separation}
K.~Singhal, R.~Rand, and M.~Amy, ``Beyond separation: Toward a specification language for modular reasoning about quantum programs,'' Programming Languages for Quantum Computing (PLanQC 2022) Poster Session, 2022, extended abstract accepted to PLanQC 2022 for the Poster Session.

\bibitem{ScaffML}
T.~Jin and J.~Zhao, ``{ScaffML}: A quantum behavioral interface specification language for scaffold,'' in \emph{2023 IEEE International Conference on Quantum Software (QSW)}, 2023, pp. 128--137.

\bibitem{Scaffold}
A.~J. Abhari, A.~Faruque, M.~J. Dousti, L.~Svec, O.~Catu, A.~Chakrabati, C.-F. Chiang, S.~Vanderwilt, J.~Black, F.~Chong, M.~Martonosi, M.~S.~K. Brown, M.~Pedram, and T.~Brun, ``Scaffold: Quantum programming language,'' Department of Computer Science, Princeton University, Tech. Rep., 2012.

\bibitem{Vellvm}
J.~Zhao, S.~Nagarakatte, M.~M. Martin, and S.~Zdancewic, ``Formalizing the {LLVM} intermediate representation for verified program transformations,'' in \emph{Proceedings of the 39th Annual ACM SIGPLAN-SIGACT Symposium on Principles of Programming Languages}, ser. POPL '12.\hskip 1em plus 0.5em minus 0.4em\relax ACM, 2012, p. 427–440.

\bibitem{QIRAlliance}
{QIR Alliance}, ``{QIR Alliance},'' \url{https://www.qir-alliance.org/}, 2023.

\bibitem{Luo23}
\BIBentryALTinterwordspacing
J.~Luo and J.~Zhao, ``Formalization of quantum intermediate representations for code safety,'' 2023. [Online]. Available: \url{arXiv:2303.14500}
\BIBentrySTDinterwordspacing

\bibitem{artifact}
\BIBentryALTinterwordspacing
M.~Lewis, S.~Soudjani, and P.~Zuliani, ``{SilVer} (artifact) v1.0.3,'' May 2024. [Online]. Available: \url{https://doi.org/10.5281/zenodo.11395797}
\BIBentrySTDinterwordspacing

\end{thebibliography}
